\documentclass[aps,prb,onecolumn,showpacs,nofootinbib,notitlepage]{revtex4-1}
\usepackage{amsmath,epsfig,amsfonts,amssymb,mathtools}
\newcommand{\be}{\begin{equation}}
\newcommand{\ee}{\end{equation}}
\newcommand{\mf}[1]{\boldsymbol{#1}}
\begin{document}
\title{Direct sampling of the self-energy with Connected Determinant Monte Carlo}
\author{Riccardo Rossi}
\affiliation{Center for Computational Quantum Physics, Flatiron Institute, 162 Fifth Avenue, New York, NY 10010, USA}
\date{\today}
\begin{abstract}
In this note, we present an efficient algorithm to sample directly the self-energy in the framework of the Connected Determinant technique. The introduction of the formalism of many-variable formal power series is essential to the proof, and more generally it is a natural mathematical tool for diagrammatic expansions.
\end{abstract}
\maketitle
\section{Introduction}
Diagrammatic Monte Carlo~\cite{DiagMC} has been proposed as an alternative to traditional Quantum Monte Carlo techniques when the latter have a sign problem, which has to be generically expected for fermionic or frustrated spin systems. Sign problem can be interpreted as a computational artifact of sampling quantities which do not have a thermodynamic limit. In traditional finite-temperature Quantum Monte Carlo methods, physical quantities are indirectly obtained as the ratio of two exponentially-large objects for which there exists sampling strategies. If these large objects happen to not be positive definite (and this is the generic case), Monte Carlo sampling is exponentially hard with system bulk size. This prevents the study of the most interesting strongly-correlated many-body systems.

Diagrammatic Monte Carlo avoids the sign problem by sampling directly physical quantities, typically Green's functions or self-energies. Having a ``sign problem'' in the sampling of physical quantities can even be advantageous for Diagrammatic Monte Carlo: this only means that the diagrammatic series will converge faster! In practice, Diagrammatic Monte Carlo, when combined with conformal-Borel resummation techniques,  is the state-of-the-art theoretical technique for the normal phase of the strongly-correlated unitary Fermi gas, where it has been benchmarked with precise cold-atom experiments~\cite{UFG,Felix_UFG}. For the Hubbard model, Diagrammatic Monte Carlo has been used to determine a large part of the zero-temperature phase diagram when one has on average less than $0.7$ particles per site~\cite{Deng}, and the approach to the pseudogap regime at finite temperature when one has about one particle per site~\cite{Ferrero_pseudo}, which is the most challenging computationally. All these results were obtained with the ``Feynman-diagrammatic'' version of the algorithm, where one samples Feynman-diagram topologies and integration variables at the same time. Recently, a new Diagrammatic Monte Carlo algorithm has been proposed\cite{Rossi}. Instead of sampling Feynman diagrams, one sums exactly over all connected Feynman-diagram topologies in an efficient way by using determinants and a recursive formula, and then one performs a Monte Carlo sampling of the internal variables of Feynman integrals. We will refer to this method as Connected Determinant Monte Carlo. The method takes advantage of the strong cancellations in fermionic Feynman-diagram topologies. In the large order limit, it has been shown to be superior to the Feynman diagram version~\cite{Rossi_EPL}, and there is also numerical evidence of this fact~\cite{Rossi, Ferrero, Kozik}. From a more fundamental point of view, when the diagrammatic series converges, the computational effort to obtain physical quantities increases only polynomially with the required precision for generic fermionic lattice systems\cite{Rossi_EPL}. Therefore, the simplest version of Diagrammatic Monte Carlo is already the state-of-the-art for weak to moderate coupling strength. The possible directions to further extend the applicability of Diagrammatic Monte Carlo are finding a convergent diagrammatic expansion in the strongly-coupled regime, and the use of efficient resummation techniques. For the first direction, one can sum-up classes of diagrams (or even use fully-self-consistent objects~\cite{Bold_DiagMC,Fedor}), one can use optimized non-interacting actions~\cite{Rossi_PRB}, or, alternatively, drastic changes of the basic degrees of freedom of the theory have been proposed~\cite{Carlstrom}. Interestingly, the resummation direction is intimately connected to the computational problem of obtaining high-order terms. The large-order behavior of the expansion can be used to detect singularities, and this information can be used to ``deform'' the complex plane of the coupling constant in order to extend the domain of convergence of the perturbative series. This technique is starting to be implemented in practical calculations in conjuction with the Connected Determinant technique to study the strongly-correlated regime of the Hubbard model~\cite{Kozik}, where it was shown to significantly extend the domain of convergence of the weak-coupling expansion.

We have seen that it is fundamental for the success of the technique in the strongly-correlated regime to be able to compute the highest number of orders possible. It is reasonable to assume that a direct sampling of the self-energy would be much more efficient than the sampling of the Green's function. For example, for high frequencies the sampling of the Green's function is dominated by the non-interacting result. Another situation where the sampling is inefficient is when the Green's function is dominated by low-order self-energy diagrams.
Following the publication of Ref.~\cite{Rossi}, there have been investigations~\cite{Ferrero_1, Kozik,Ferrero} directed to understand what is the most efficient way to implement the direct sampling of the self-energy. In Ref.~\cite{Ferrero} it was found that the algorithm for the direct sampling of the self-energy sketched in Ref.~\cite{Rossi} is not optimal, and two modifications have been proposed. In Ref.~\cite{Kozik} another modification was proposed to directly sample the (completely symmetrized) self-energy in the momentum representation.

In this note, we prove that the self-energy can be directly sampled in the momentum representation with the same algorithm of the Green's function. In particular, we compute the same quantity as in Ref.~\cite{Kozik}, with a smaller computational cost. Moreover, the proof is completely general and it applies to every possible model and diagrammatic expansion. The result is obtained by introducing the mathematical formalism of many-variable formal power series. While the derivation of the Connected Determinant algorithm for the Green's function can be obtained by an intuitive Feynman-diagram picture, as it was done in Ref.~\cite{Rossi}, the Feynman-diagram interpretation of the recursive formula for the self-energy is less straightforward. This shows the superior power of purely-algebraic methods compared to intuitive Feynman diagrams representations.

This note is organized in two parts: In Section II we introduce and motivate the formalism of many-variable formal power series. In Section III we present, as a first non-trivial application, the direct sampling of the self-energy. 
\section{Diagrammatic expansions and many-variable formal power series}
In this section we introduce a general and powerful framework to formalize diagrammatic expansions. With this formalism the recursive formula of Ref.~\cite{Rossi} is the {\it definition} of the division between many-variable formal power series, and no Feynman diagram needs to be introduced.
\subsection{Motivation}
In order to motivate the formalism, let us start with an example. We would like to stress that the formalism is completely general, it can be applied to any model, diagrammatic expansion, or correlation function. Let us consider the  Hubbard model, which describes two species of fermions hopping in a hypercubic lattice experiencing onsite repulsion. The hamiltonian operator $\hat{H}$ is
\be
\hat{H}:=-\sum_{\mf{r}\in \Lambda}\sum_{j=1}^d\sum_{\sigma\in\{\downarrow,\uparrow\}}\left(\hat{\psi}_\sigma^\dagger(\mf{r}) \hat{\psi}_\sigma(\mf{r+e}_j)+\hat{\psi}_\sigma^\dagger(\mf{r+e}_j)  \hat{\psi}_\sigma(\mf{r})\right)+U\sum_{\mf{r}\in \Lambda}(\hat{\psi}_\uparrow^\dagger \hat{\psi}_\downarrow^\dagger \hat{\psi}_\downarrow \hat{\psi}_\uparrow)(\mf{r})
\ee
where $U\in\mathbb{C}$, $(\mf{e}_j)_k=\delta_{jk}$, $\Lambda =(\mathbb{Z}/(L\mathbb{Z}))^d$, $L<\infty$, and $\hat{\psi}_\sigma$ are destruction fermionic operators, defined by the anticommutations relations $\{\hat{\psi}_{\sigma_1}(\mf{r}_1),\hat{\psi}_{\sigma_2}(\mf{r}_2)\}:=0$, $\{\hat{\psi}_{\sigma_1}^\dagger(\mf{r}_1),\hat{\psi}_{\sigma_2}(\mf{r}_2)\}:=\delta_{\sigma_1,\sigma_2}\,\delta_{\mf{r}_1,\mf{r}_2}$. In order to be able to select the number of particles, we add a chemical potential term to the hamiltonian $\hat{H}':=\hat{H}-\mu\sum_{\sigma}\sum_{\mf{r}\in\Lambda}(\hat{\psi}_\sigma^\dagger\hat{\psi}_\sigma)(\mf{r})$. In order to compute the properties of this many-body system at thermal equilibrium, is it useful to consider the space+imaginary-time operators destruction operators
\be
\hat{\psi}_\sigma(\mf{r},\tau):=e^{\tau\hat{H}'}\,\hat{\psi}_\sigma(\mf{r})\,e^{-\tau\hat{H}'}
\ee
where $\tau\in\mathbb{R}$ is the ``imaginary time''.
 We introduce the Green's function $G$ of the Hubbard model:
\be
G_{(\boldsymbol{r},\tau)}(U):= - \frac{\text{Tr}\left\{e^{-\beta\hat{H}'}\;T_{\text{ord}} \left[\hat{\psi}_{\uparrow}(\mf{r},\tau) \,\hat{\psi}_{\uparrow}^\dagger(\mf{0},0)\right]\right\}}{\text{Tr}\;e^{-\beta \hat{H}'}}
\ee
where $\mf{r}\in\Lambda$,  $U\in\mathbb{C}$, $\tau\in(-\beta,\beta)\setminus \{0\}$, $0<\beta<\infty$, $\beta$ is the inverse temperature, and the time-ordering operator is defined by 
\be
T_{\text{ord}} \left[\hat{\psi}_{\uparrow}(\mf{r},|\tau|) \,\hat{\psi}_{\uparrow}^\dagger(\mf{0},0)\right]=\hat{\psi}_{\uparrow}(\mf{r},|\tau|) \,\hat{\psi}_{\uparrow}^\dagger(\mf{0},0)
\ee
\be
T_{\text{ord}} \left[\hat{\psi}_{\uparrow}(\mf{r},-|\tau|) \,\hat{\psi}_{\uparrow}^\dagger(\mf{0},0)\right]= -\hat{\psi}_{\uparrow}^\dagger(\mf{0},0)\,\hat{\psi}_{\uparrow}(\mf{r},-|\tau|)
\ee
The Green's function for $U=0$, which we call $G^{(0)}$, can be computed exactly. More generally, for $U\neq 0$,  the Green's function can be computed from a convergent series in powers of $U$ for $|U|<R$:
\be
G_{(\mf{r},\tau)}(U)=\sum_{n=0}^\infty \frac{U^n}{n!}\; G_{(\mf{r},\tau)}^{(n)},\qquad |U|<R
\ee
where $R>0$ is the radius of convergence. It can be proven mathematically that quite generically for fermionic lattice models this is the standard situation~\cite{Benfatto}, and there is also extensive numerical evidence~\cite{Ferrero_pseudo, Rossi}.
It is well known (see, e.g., Ref.~\cite{Abrikosov}) that $G_{(\mf{r},\tau)}^{(n)}$ for $n\ge 1$ can be expressed as a space-time sum-integral over the positions of the interaction vertices $(\mf{r}_1,\tau_1),(\mf{r}_2,\tau_2),\dots,(\mf{r}_{n},\tau_{n})$:
\be
G_{(\mf{r},\tau)}^{(n)} = \sum_{\mf{r}_1,\dots,\mf{r}_{n}\in \Lambda}\int_{[0,\beta]^n}d\tau_1\dots d\tau_{n}\;G_{(\mf{r},\tau)}^{\text{Feyn}}((\mf{r}_1,\tau_1),\dots,(\mf{r}_n,\tau_n))
\ee
where $G_{(\mf{r},\tau)}^{\text{Feyn}}((\mf{r}_1,\tau_1),\dots,(\mf{r}_n,\tau_n))$ is defined as the sum of all {\it connected} Feynman diagram topologies $\mathcal{F}_n$ for the Green's function with $n$ interaction vertices at fixed space-time positions
\be
G_{(\mf{r},\tau)}^{\text{Feyn}}((\mf{r}_1,\tau_1),\dots,(\mf{r}_n,\tau_n)):=\sum_{\mathcal{T}\in\mathcal{F}_n}\,\mathcal{D}_{(\mf{r},\tau)}(\mathcal{T}|(\mf{r}_1,\tau_1),\dots,(\mf{r}_n,\tau_n))
\ee
$|\mathcal{F}_n|$ is of the order of $n!$ for general two-body interactions. We extend the definition of the Green's function to be a formal {\it functional} of a space-time complex field $\xi(\mf{r},\tau)$:
\be
G_{(\mf{r},\tau)}[\xi]:=\sum_{n=0}^\infty\frac{1}{n!}\sum_{\mf{r}_1,\dots,\mf{r}_{n}\in \Lambda}\int_{[0,\beta]^n}d\tau_1\dots d\tau_{n}\;\left(\prod_{j=1}^n \xi(\mf{r}_j,\tau_j)\right) G_{(\mf{r},\tau)}^{\text{Feyn}}((\mf{r}_1,\tau_1),\dots,(\mf{r}_n,\tau_n))
\ee
For $\xi(\mf{r},\tau)=U$, one has formally $G_{(\mf{r},\tau)}[\xi]=G_{(\mf{r},\tau)}(U)$. $G_{(\mf{r},\tau)}[\xi]$ can be interpreted as the Green's function in a space-time dependent coupling constant $\xi(\mf{r},\tau)$. We remark that only the symmetric part of $G^{\text{Feyn}}$ contributes to the functional. Therefore, we define
\be
G_{(\mf{r},\tau)}(\{(\mf{r}_1,\tau_1),\dots,(\mf{r}_n,\tau_n)\}):=\frac{1}{n!}\sum_{\sigma\in S_n}G_{(\mf{r},\tau)}^{\text{Feyn}}((\mf{r}_{\sigma(1)},\tau_{\sigma(1)}),\dots,(\mf{r}_{\sigma(n)},\tau_{\sigma(n)}))
\ee
where $\sigma\in S_n$ is a permutation of $n$ objects. Note that in the left hand side of the previous equation we use the set notation for the vertices as the order of them is not important. Another important remark to make is that $G_{(\mf{r},\tau)}(\{(\mf{r}_1,\tau_1),\dots,(\mf{r}_n,\tau_n)\})$ can be written as the sum of $\sim (n!)^2$ Feynman diagrams for two-body interactions (for $k$-body interactions the number is $\sim(n!)^k$). It would be essentially hopeless to compute this object for large $n$ using the brute-force Feynman-diagram definition, while the Connected Determinant technique allows to compute this object with a number of arithmetic operations always equal to $3^n$ (see the next subsection for a proof). We can then write
\be
G_{(\mf{r},\tau)}[\xi]:=\sum_{n=0}^\infty\frac{1}{n!}\sum_{\mf{r}_1,\dots,\mf{r}_{n}\in \Lambda}\int_{[0,\beta]^n}d\tau_1\dots d\tau_{n}\;\left(\prod_{j=1}^n \xi(\mf{r}_j,\tau_j)\right) G_{(\mf{r},\tau)}(\{(\mf{r}_1,\tau_1),\dots,(\mf{r}_n,\tau_n)\})
\ee
 In other terms, $G_{(\mf{r},\tau)}(\{(\mf{r}_1,\tau_1),\dots,(\mf{r}_n,\tau_n)\})$ is the coefficient of $\prod_{j=1}^n \xi(\mf{r}_j,\tau_j)$ in the functional expansion:
\be
\left.\frac{\delta^n G_{(\mf{r},\tau)}[\xi]}{\delta \xi(\mf{r}_1,\tau_1)\dots \delta\xi(\mf{r}_n,\tau_n)}\right|_{\xi=0}=G_{(\mf{r},\tau)}(\{(\mf{r}_1,\tau_1),\dots,(\mf{r}_n,\tau_n)\})
\ee
The previous equation shows that $G_{(\mf{r},\tau)}(\{(\mf{r}_1,\tau_1),\dots,(\mf{r}_n,\tau_n)\})$ can be interpreted a non-linear high-order response function of the non-interacting system to the change of the coupling constant. We write
\be
G_{(\mf{r},\tau)}(U)=\frac{A_{(\mf{r},\tau)}(U)}{Z(U)}
\ee
where $A_{(\mf{r},\tau)}(U):=- \text{Tr}\left\{e^{-\beta\hat{H}'}\;T_{\text{ord}} \left[\hat{\psi}_{\uparrow}(\mf{r},\tau) \,\hat{\psi}_{\uparrow}^\dagger(\mf{0},0)\right]\right\}/z_0$, $Z(U):=\text{Tr}\;e^{-\beta \hat{H}'}/z_0$, and $z_0:= \text{Tr}\;e^{-\beta \hat{H}'}|_{U=0}$. Reasoning as before, we can extend the definition of $A$ and $Z$ to be functionals of a space-time dependent coupling constant $\xi(\mf{r},\tau)$, to obtain $A_{(\mf{r},\tau)}[\xi]$ and $Z[\xi]$.
Then, we can write the  Green's functional $G_{(\mf{r},\tau)}[\xi]$ as the ratio of two other functionals:
\be\label{Green_functional_ratio}
G_{(\mf{r},\tau)}[\xi]=\frac{A_{(\mf{r},\tau)}[\xi]}{Z[\xi]}
\ee
where the coefficients of the expansions for $A_{(\mf{r},\tau)}[\xi]$ and $Z[\xi]$ can be efficiently computed with Wick's theorem~\cite{Schwinger}:
\be
A_{(\mf{r},\tau)}(\{(\mf{r}_1,\tau_1),\dots,(\mf{r}_n,\tau_n)\})=(-1)^n\det \mathbb{A}_{(\mf{r},\tau)}\;\det \mathbb{Z}
\ee 
\be
Z(\{(\mf{r}_1,\tau_1),\dots,(\mf{r}_n,\tau_n)\})=(-1)^n\left(\det \mathbb{Z}\right)^2
\ee
where $\mathbb{A}_{(\mf{r},\tau)}$ and $\mathbb{Z}$ are respectively $(n+1)\times (n+1)$ and $n\times n$ matrices defined by 
\be\label{def_A}
(\mathbb{A}_{(\mf{r},\tau)})_{ab}=(\mathbb{Z})_{ab}=G^{(0)}_{(\mf{r}_a-\mf{r}_b,\tau_a-\tau_b-0^+)},\qquad a,b\in\{1,\dots,n\}
\ee
\be
(\mathbb{A}_{(\mf{r},\tau)})_{0b}=G^{(0)}_{(\mf{r}-\mf{r}_b,\tau-\tau_b-0^+)},\qquad b\in\{1,\dots,n\}
\ee
\be
(\mathcal{A}_{(\mf{r},\tau)})_{a0}=G^{(0)}_{(\mf{r}_a,\tau_a-0^+)},\qquad a\in\{1,\dots,n\}
\ee
and $(\mathbb{A}_{(\mf{r},\tau)})_{00}=G^{(0)}_{(\mf{r},\tau)}$. This is the basis of the determinant diagrammatic Monte Carlo algorithm~\cite{Bourovski,Rubtsov}, and the related interaction-expansion continuos-time Monte Carlo~\cite{CT-INT}. In the Feynman-diagram interpretation, $\\A_{(\mf{r},\tau)}(\{(\mf{r}_1,\tau_1),\dots,(\mf{r}_n,\tau_n)\})$ is the sum of {\it{all}} symmetrized Feynman-diagram topologies (connected and disconnected) of Green's function. Similarly, $Z(\{(\mf{r}_1,\tau_1),\dots,(\mf{r}_n,\tau_n)\})$ is the sum of connected and disconnected symmetrized Feynman-diagram topologies without external legs. The computational effort to compute these determinants increases polynomially with the order of the expansion for $A$ and $Z$. For a given set $\{(\mf{r}_1,\tau_1),\dots,(\mf{r}_n,\tau_n)\}$, one can compute $A_{(\mf{r},\tau)}(\{(\mf{r}_1,\tau_1),\dots,(\mf{r}_n,\tau_n)\})$ and $Z(\{(\mf{r}_1,\tau_1),\dots,(\mf{r}_n,\tau_n)\})$ in $O(n^3)$ arithmetic operations.

However, the object which has a physical importance is the Green's function, and we would like to find a direct algorithm to compute it without first computing $A$ and $Z$ and taking the ratio between these two at the end of the calculation. There is also a much more important  reason to not compute $A$ and $Z$: they are macroscopically large objects, they increase exponentially with the system bulk size. This means that if we have a system with sign problem (which is the case, for instance, of the repulsive Hubbard model away from half filling), it would be extremely challenging to extrapolate to the infinite-size limit. The Green's function can be defined directly in the thermodynamic limit where the linear system size $L$ goes to infinity, and so is  every coefficient of the functional $G_{(\mf{r},\tau)}[\xi]$. Reasoning in terms of many-variable formal power series, we can obtain the Green's function as the ratio of two power series, as shown by Equation~\eqref{Green_functional_ratio}.  Therefore, in the next subsection we develop the algebraic theory of many-variable formal power series.

\subsection{Many-variable formal power series}
We introduce the commuting variables $\xi_v$, indexed by a discrete label $v$ belonging to a set $\mathcal{I}$, $v\in \mathcal{I}$, $|\mathcal{I}|<\infty$. We suppose that $\mathcal{I}$ has an order relation, that is for $v_1,v_2\in\mathcal{I}$, one has $v_1\le v_2$ or $v_2\le v_1$. The continuum case can be obtained as a limiting case of the discrete case, but the latter theory is more general. A many-variable formal power series $f[\xi]$ is defined by its coefficients $f(\{v_1,\dots,v_n\})\in\mathbb{C}$, $v_k\in \mathcal{I}$, $k\in\{1,\dots,n\}$, $n\in\mathbb{N}_0$. We write formally
\be
f[\xi]=\sum_{n=0}^\infty\, \sum_{v_1\le v_2\le \dots \le v_n,\, v_k\in \mathcal{I}}\left(\prod_{j=1}^n\xi_{v_j}\right)\,f(\{v_1,\dots,v_n\})
\ee
In the previous formula, $\xi_v$ is just a commuting symbol. No numerical value needs to be associated to it at this stage. We introduce a useful notation: for $V=\{v_1,\dots,v_n\}$, we define
\be
\xi^V:=\prod_{j=1}^n\xi_{v_j}
\ee
We can then write
\be
f[\xi] =\sum_{n=0}^\infty \sum_{|V|=n}\xi^V\;f(V)=:\sum_{V} \xi^V\;f(V)
\ee
where the sum goes over all multisets built with $\mathcal{I}$. We also remark that if $V$ and $W$ are such multisets of $\mathcal{I}$, one has
\be\label{xi_times_xi}
\xi^V\,\xi^W=\xi^{V\cup W}
\ee
where $V\cup W$ is the union between multisets. We are now ready to introduce the fundamental algebraic operations between many-variable formal power series. The addition is simply:
\be
f[\xi] +g[\xi] = \sum_V \xi^V(f(V)+g(V))
\ee
This means that the zero element for the addition is the series with all coefficients equal to zero: $f[\xi]=0$ is equivalent to $f(V)=0$ for all $V$ multisets of $\mathcal{I}$.
The multiplication, using Equation~\eqref{xi_times_xi}, is
\be
f[\xi]\,g[\xi]=\sum_V \xi^V\sum_{S\subseteq V} f(V\setminus S) \,g(S)
\ee
This is also known as the Cauchy product for one-variable formal power series (which corresponds to the case $|\mathcal{I}|=1$).  The number of multiplications needed to obtain the coefficient of $\xi^V$ of a multiplication of two many-variable formal power series is $\prod_{j=1}^m (r_j+1)$ if the multiset $V$ consists of $r_j$ repetitions of the element $v_j$, $j\in\{1,\dots,m\}$, i.e. $V=\cup_{j=1}^m \cup_{k=1}^{r_j} \{v_j\}$, and $v_j\neq v_l$ for $j\neq l$. In the particular case where all the elements are the same ($r_1=|V|$, $m=1$), the number of operations is $|V|+1$. If all elements are different ($r_j=1$, $m=|V|$), the number of operations is $2^{|V|}$. 

We can also define the division between two many-variable power series $h[\xi]=f[\xi]/g[\xi]$ as a solution of this equation
\be
h[\xi]\,g[\xi] =f[\xi]
\ee
which exists and it is unique when $g[0]=g(\emptyset)\neq 0$. In this case, we can compute $h[\xi]$ with a recursive formula:
\be\label{division_formula}
h(V)=\frac{f(V)}{g(\emptyset)}-\sum_{S\subsetneq V} h(S)\;\frac{g(V\setminus S)}{g(\emptyset)}
\ee
Let us count the number of multiplications needed to obtain $h(V)$. Without loss of generality, we consider the case $g(\emptyset)=1$. First of all, we introduce as before the number of repetitions in the set $V$, denoted by $r_j$, $j\in\{1,\dots,m\}$ (i.e. $V=\cup_{j=1}^m \cup_{k=1}^{r_j} \{v_j\}$, and $v_j\neq v_l$ for $j\neq l$). Let $W\subseteq V$. $W$ is identified by the number of repetitions $x_1$ of the first element of $V$, $0\le x_1\le r_1$, the number of repetitions $x_2$ of the second element of $V$, $0\le x_2\le r_2$, and so on. Suppose now that we have computed $h(S)$ for all $S\subsetneq W$. In order to compute $h(W)=f(W)-\sum_{S\subsetneq W} h(S)\,g(W\setminus S)$, we need a number of multiplications which is equal to the number of proper subsets of $W$, which is $\prod_{j=1}^m(x_j+1)-1$. We perform this operation for every $W\subseteq V$, starting from $W=\emptyset$ (there is nothing to do in this case). The total number of multiplications for computing $h(V)$ is then
\be
\sum_{x_1=0}^{r_1}\dots\sum_{x_m=0}^{r_m}\left[\prod_{j=1}^m(x_j+1)-1\right]=\left(\prod_{j=1}^m(r_j+1)\right)\left(\prod_{j=1}^m\frac{r_j+2}{2}-1\right)
\ee
Let us consider the case of a one-variable power series, where $m=1$ and $r_1=|V|$. We then see that the computational cost is $O(|V|^2)$. In this article we are mainly interested in the case where $m=|V|$. In this case, the computational effort is $O(3^{|V|})$.

Let us briefly consider two important applications of this formalism, the computation of the Green's function and of the free energy, that were presented in Ref.~\cite{Rossi} using an intuitive graphical derivation. The Green's function can be computed as the division between two many-variable formal power series, see Equation~\eqref{Green_functional_ratio}. We will skecth the computation of the free-energy, that we define here as the logarithm of some partition function:
\be
p[\xi]=\log Z[\xi]
\ee
Taking the Euler's derivative, we have
\be
E[\xi]:=\sum_v \xi_v \frac{\partial}{\partial \xi_v} p[\xi]=\sum_V \xi^V \,|V|\,p(V)=\frac{\sum_V \xi^V\,|V|\, Z(V)}{Z[\xi]}=:\frac{Z_1[\xi]}{Z[\xi]}
\ee
Therefore, the free-energy coefficients $p(V)$ can be computed in the same way as the Green's function.
\section{Application: recursive formula for the self-energy}
We are now ready to present the derivation of the recursive formula for the self-energy directly in the momentum representation for the external points. We perform a Fourier transform on the Green's function:
\be
\mathcal{G}_{(\mf{k},\omega)}(U):=\sum_{\mf{r}\in \Lambda} \int_0^\beta d\tau\;e^{-i\mf{k}\cdot\mf{r}+i\omega \tau}\;G_{(\mf{r},\tau)}(U)
\ee
where $\mf{k}\in\mathbb{R}^d$, and $\omega\in\mathbb{R}$. We have now all the elements to introduce the self-energy $\Sigma$ from the Dyson equation
\be
\Sigma_{(\mf{k},\omega)}(U):=[\mathcal{G}_{(\mf{k},\omega)}^{(0)}]^{-1}-[\mathcal{G}_{(\mf{k},\omega)}(U)]^{-1}
\ee
As we have done for the Green's function, we extend the definition of the self-energy to allow for a space-time dependent interaction
\be
\Sigma_{(\mf{k},\omega)}[\xi]:=\sum_{n=0}^\infty\frac{1}{n!}\sum_{\mf{r}_1,\dots,\mf{r}_{n}\in \Lambda}\int_{[0,\beta]^n}d\tau_1\dots d\tau_{n}\;\left(\prod_{j=1}^n \xi(\mf{r}_j,\tau_j)\right) \Sigma_{(\mf{k},\omega)}(\{(\mf{r}_1,\tau_1),\dots,(\mf{r}_n,\tau_n)\})
\ee
where, as before, $\Sigma_{(\mf{k},\omega)}(\{(\mf{r}_1,\tau_1),\dots,(\mf{r}_n,\tau_n)\})$ can be interpreted as the sum of all self-energy (therefore irreducible) Feynman diagrams with external momentum equal to $(\mf{k},\omega)$ and with interaction vertices at space-time positions $\{(\mf{r}_1,\tau_1),\dots,(\mf{r}_n,\tau_n)\}$ (we remind that the position of the interaction vertices is automatically symmetrized, therefore are $\sim (n!)^2$ diagrams for two-body interactions). We can express the self-energy functional directly in terms of the Green's functional
\be\label{sigma_ratio}
\Sigma_{(\mf{k},\omega)}[\xi]=[\mathcal{G}_{(\mf{k},\omega)}^{(0)}]^{-1}-[\mathcal{G}_{(\mf{k},\omega)}[\xi]]^{-1}=[\mathcal{G}_{(\mf{k},\omega)}^{(0)}]^{-1}-\frac{Z[\xi]}{\mathcal{A}_{(\mf{k},\omega)}[\xi]}
\ee
where 
\be
\mathcal{A}_{(\mf{k},\omega)}[\xi]=\sum_V \xi^V\, \mathcal{A}_{(\mf{k},\omega)}(V),\qquad \mathcal{A}_{(\mf{k},\omega)}(V):=\sum_{\mf{r}\in \Lambda} \int_0^\beta d\tau\;e^{-i\mf{k}\cdot\mf{r}+i\omega \tau}\;A_{(\mf{r},\tau)}(V)
\ee
$\mathcal{A}_{(\mf{k},\omega)}(V)$ can be expressed in terms of determinants (see below for explicit expressions for the Hubbard model). One has for $\Sigma_{(\mf{k},\omega)}(\emptyset)=0$. Applying the division formula~\eqref{division_formula} to Equation~\eqref{sigma_ratio}, one has (for $V\neq \emptyset$)
\be\label{recursive_self}
\Sigma_{(\mf{k},\omega)}(V)=\frac{1}{\mathcal{G}^{(0)}_{(\mf{k},\omega)}}\,\frac{\mathcal{A}_{(\mf{k},\omega)}(V)}{\mathcal{A}_{(\mf{k},\omega)}(\emptyset)}-\frac{Z(V)}{\mathcal{A}_{(\mf{k},\omega)}(\emptyset)}-\sum_{S\subsetneq V} \Sigma_{(\mf{k},\omega)}(S)\;\frac{\mathcal{A}_{(\mf{k},\omega)}(V\setminus S)}{\mathcal{A}_{(\mf{k},\omega)}(\emptyset)}
\ee
Therefore, the computational cost is $O(3^{|V|})$. For concreteness, we give explicit expressions for $\mathcal{A}_{(\mf{k},\omega)}$ for the Hubbard model:
\be
\mathcal{A}_{(\mf{k},\omega)}(\emptyset)=\mathcal{G}_{(\mf{k},\omega)}^{(0)},\qquad \frac{\mathcal{A}_{(\mf{k},\omega)}(V)}{\mathcal{A}_{(\mf{k},\omega)}(\emptyset)}=(-1)^n\,\det \,\mathbb{B}_{(\mf{k},\omega)}\;\det \mathbb{Z}
\ee
where $\mathbb{Z}$ is defined in Equation~\eqref{def_A},  $\mathbb{B}_{(\mf{k},\omega)}$ is a $(n+1)\times (n+1)$ matrix defined by $(\mathbb{B}_{(\mf{k},\omega)})_{ab}=(\mathbb{A}_{(\mf{r},\tau)})_{ab}$ for $a\in\{1,\dots,n\}$, $b\in \{0,\dots,n\}$ (see Equation~\eqref{def_A} for the definition) and 
\be
(\mathbb{B}_{(\mf{k},\omega)})_{0b}= e^{-i\mf{k}\cdot \mf{r}_b+i\omega \tau_b},\qquad b\in\{1,\dots,n\}
\ee
and $(\mathbb{B}_{(\mf{k},\omega)})_{00}=1$.
\section{Conclusion}
In conclusion, we have presented an efficient algorithm for the direct sampling of the self-energy with the Connected Determinant method. From a computational point of view, the algorithm is identical to the one of the Green's function presented in Ref.~\cite{Rossi}, and it is an improvement over the self-energy algorithm presented therein, and developed further in Ref.~\cite{Kozik, Ferrero}. It is interesting to remark that in order to obtain the result it is essential to use the elegant formalism of many-variable formal power series, which we have introduced in this note. This proves the superiority of this algebraic method over the graphical Feynman-diagram description.

The formalism of many-variable formal power series could provide other interesting applications. For example, it is possible to smoothly interpolate between Diagrammatic Monte Carlo and traditional Quantum Monte Carlo techniques using this formalism, with the hope that the ``hybridized'' Monte Carlo shows superior applicability of both techniques. For instance, one could use determinant Quantum Monte Carlo for local correlations and the division formula for non-local ones by considering a space-dependent coupling constant (instead of a space-time dependent coupling constant). We have presented the discrete theory of many-variable formal power series with this application in mind. Another application is the reduction of the variance in the Connected Determinant Monte Carlo sampling, which is obtained in essence by summing over spacetime vertices positions before applying the recursive formula.
\section{Acknowledgments}
I acknowledge useful discussions with Michel Ferrero, Fedor Simkovic, Hugo Strand, Kris Van Houcke, and F\'elix Werner. I would like to thank Evgeny Kozik for useful discussions and for spotting a critical typo. I would also like to thank all the partecipants of the Diagrammatic Monte Carlo workshop of June 2017, held at the Flatiron Institute, New York. The Flatiron Institute is a division of the Simons Foundation. I acknowledge support from the Simons Foundation's Many Electron Collaboration.
\bibliography{biblio} 
\end{document}